\documentclass{article}
\usepackage{arxiv}
\usepackage[utf8]{inputenc} 
\usepackage[T1]{fontenc}    
\usepackage{hyperref}       
\usepackage{algorithm,algpseudocode}
\usepackage{amsmath}
\usepackage{amsfonts}
\usepackage{amssymb}

\usepackage{graphicx}
\graphicspath{{./fig/}}
\usepackage{listings}
\usepackage{xcolor}
\usepackage{caption}
\usepackage{subcaption}
\usepackage{multirow}
\newcommand{\squeezeupless}{\vspace{-2mm}}
\newcommand{\squeezeup}{\vspace{-4mm}}

\usepackage{cite}
\usepackage{amsmath,amssymb,amsfonts}
\usepackage{graphicx}
\usepackage{textcomp}
\usepackage{xcolor}
\def\BibTeX{{\rm B\kern-.05em{\sc i\kern-.025em b}\kern-.08em
    T\kern-.1667em\lower.7ex\hbox{E}\kern-.125emX}}
\begin{document}

\title{Exploring Memory Persistency Models for GPUs}

\author{%
  \makebox[.45\linewidth]{Zhen Lin}\\
  Department of Electrical and Computer Engineering\\
  North Carolina State University\\
  Raleigh, NC \\
  \texttt{zlin4@ncsu.edu} \\
  \and 
  \makebox[.45\linewidth]{\bf{Mohammad Alshboul}}\\
  Department of Electrical and Computer Engineering\\
  North Carolina State University\\
  Raleigh, NC \\
  \texttt{maalshbo@ncsu.edu} \\
  \and 
  \makebox[.45\linewidth]{\bf{Yan Solihin}}\\
  Department of Computer Science\\
  University of Central Florida\\
  Orlando, FL \\
  \texttt{yan.solihin@ucf.edu} \\
  \and 
  \makebox[.45\linewidth]{\bf{Huiyang Zhou}}\\
  Department of Electrical and Computer Engineering\\
  North Carolina State University\\
  Raleigh, NC \\
  \texttt{hzhou@ncsu.edu} \\
}

\maketitle

\begin{abstract}

Given its high integration density, high speed, byte addressability, and low standby power, non-volatile or persistent memory is expected to supplement/replace DRAM as main memory. Through persistency programming models (which define durability ordering of stores) and durable transaction constructs, the programmer can provide recoverable data structure (RDS) which allows programs to recover to a consistent state after a failure. While persistency models have been well studied for CPUs, they have been neglected for graphics processing units (GPUs). Considering the importance of GPUs as a dominant accelerator for high performance computing, we investigate persistency models for GPUs. 

GPU applications exhibit substantial differences with CPUs applications, hence in this paper we adapt, re-architect, and optimize CPU persistency models for GPUs. We design a pragma-based compiler scheme to express persistency models for GPUs. We identify that the thread hierarchy in GPUs offers intuitive scopes to form epochs and durable transactions. We find that undo logging produces significant performance overheads. We propose to use idempotency analysis to reduce both logging frequency and the size of logs. Through both real-system and simulation evaluations, we show low overheads of our proposed architecture support. 

\end{abstract}


\section{Introduction}
\label{sec:intro}

Non-volatile memory (NVM) or Persistent Memory (PM) is here~\cite{optane} and is expected to supplement/replace DRAM as main memory due to its high integration density, comparably high read speed, byte addressability, and low leakage power. An example PM is Intel Optane DC Persistent Memory~\cite{optanedc}, which is a DDR4-connected device supporting 3TB/socket main memory. The non-volatility makes it possible to host data persistently in main memory, blurring the boundary between main memory and storage, thereby challenging the classical computer system design. 

Persistent data storage in main memory provides an opportunity to achieve {\em recoverable data structures} (RDS), which allows programs to recover from crashes just by using data in main memory instead of a checkpoint. Recent research~\cite{elnawawy17, lazypersistency} showed that by relying on RDS instead of checkpoints, highly significant performance and write endurance improvements can be obtained. Achieving RDS requires {\em persistency models} along with instruction support, and a crash recovery technique such as logging. Various memory persistency models have been proposed for CPUs, defining the order in which stores become durable in main memory, often in relation to ordering defined in memory consistency models regarding when stores become visible to other threads in a parallel program~\cite{pelley15}. In addition to store durability ordering, ensuring a consistent data at any given point in time is required, typically supported through durable transactions and their associated logging mechanisms~\cite{intel, kolli16, volos11}. 


While persistency models in CPUs have been well explored, they have been neglected in GPUs. We envision GPU will make use of persistent memory in the near future for the following key reasons besides the increased memory capacity over volatile DRAM. First, in current systems, persistent data is kept in files, and must be read to build data structures at process start, written to files at process termination. The conversion between persistent and temporary is expensive and unnecessary with NVM. One such use case is in-memory databases, especially the GPU accelerated ones including Mega-KV \cite{megakv15}, GPU B-Tree \cite{btree19}, Kinetica \cite{kinetica}, etc. Second, long-running GPU applications, including training deep neural networks, computing proof of work in blockchain applications, scientific computation using iterative approaches, etc., would benefit from fault tolerance with RDS. Having recoverable persistent data in NVM allows the processor to recover from soft errors. This relegates system checkpointing for more serious faults, hence the checkpointing frequency can be reduced \cite{elnawawy17}. Third, in fusion-like architecture, CPU and GPU share PM. Therefore, we argue that GPU needs to support memory persistency models. 
 
\begin{figure}
\centering
\includegraphics[width=.6\linewidth, , trim=2 2 2 2, clip]{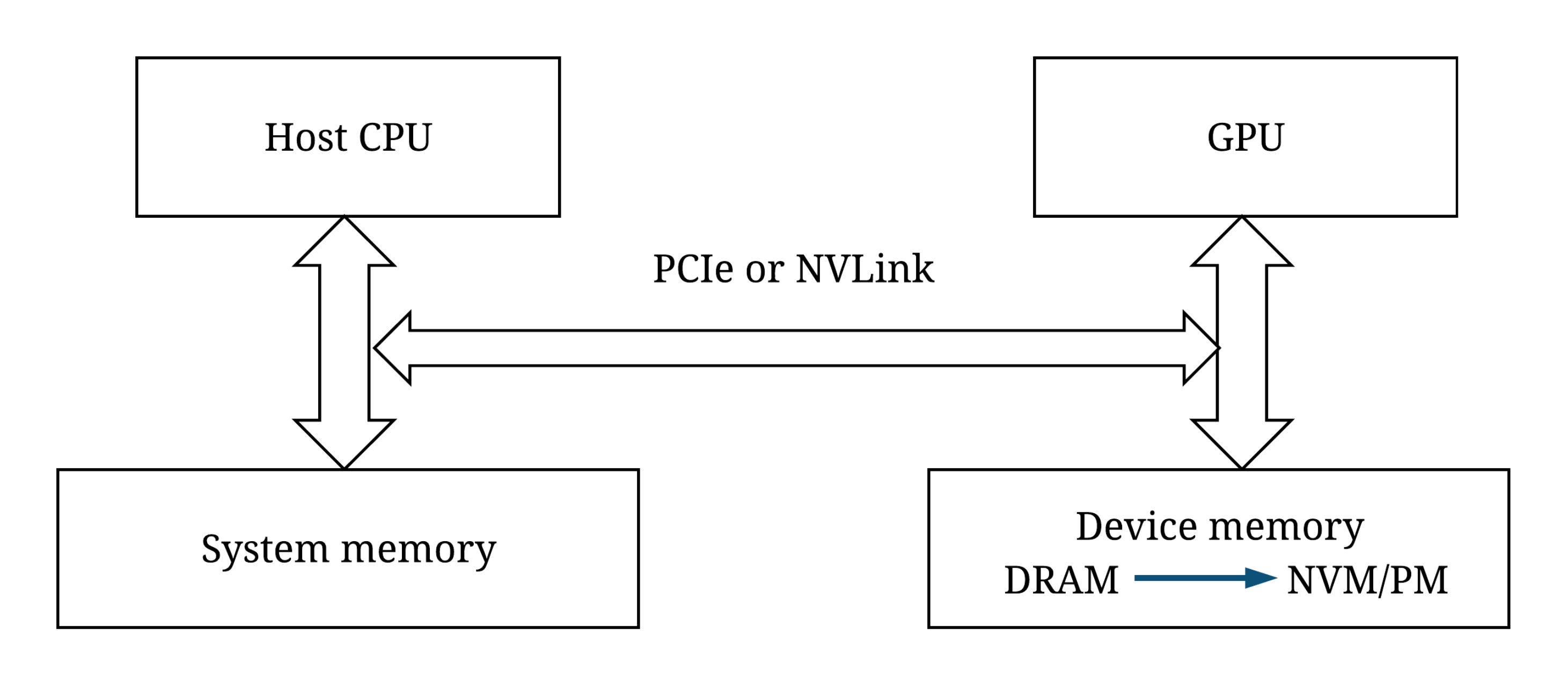}
\caption{The system architecture with a discrete GPU, for which NVM/PM replaces DRAM as device memory.}
\squeezeup
\label{fig:system}
\end{figure}

In our study, we assume discrete GPU systems shown in Figure \ref{fig:system}, since discrete GPUs have high memory bandwidth and are most commonly used in HPC (High-Performance Computing) systems although the same support can be adopted for fusion-like  architectures. Recent GPUs support both unified and non-unified memory \cite{sakharnykh16}. With non-unified memory, the programmer needs to explicitly copy the data between host-side system memory and device memory while with unified memory, the memory pages can be migrated on-demand, reducing the programmer's complexity. In this work, we consider both unified and non-unified memory models and assume that the persistency of the host-side system memory is properly handled with existing approaches~\cite{pelley15, intel, kolli16, pelley14}.


CPU memory persistency needs to be re-architected for GPUs, because of the following key differences: {\bf Workloads}: CPU benchmarks utilizing PM, such as Whisper~\cite{whisper}, mainly focus on database applications. In comparison, long-running GPU tasks also include scientific computations, deep neural network training, large graph processing, and blockchain mining. They exhibit different characteristics and execution behavior. {\bf Bandwidth vs. latency}: memory-intensive kernels in GPUs are  typically bandwidth- (instead of latency-) sensitive. Latency-oriented optimizations for CPUs such as durable write-pending queues (WPQs) at the memory controller have limited impact on GPUs. Furthermore, creating/updating logs introduce additional write traffic/bandwidth, which affect GPUs much more than CPUs. {\bf Multiple Memory Partitions/Controllers}: a GPU is equipped with multiple memory partitions and memory controllers for high bandwidth. The pcommit instruction, which makes pending writes durable, needs to be broadcasted to all MCs to flush all their WPQs. 
{\bf Scratchpad memory}: a typical server/desktop GPU memory hierarchy has scratchpad memory (aka shared memory), which needs to be considered for GPU memory persistency. 

{\em In this paper, we adapt, re-architect, and optimize CPU persistency models for GPUs.} The paper makes the following contributions: 

First, to provide both simplicity and flexibility to use the persistency models, we propose {\sf pragmas} for programmers to specify their choice of persistency model and code region, allowing the compiler to automatically generate GPU persistency code. 

Second, we propose GPU-friendly implementation of strict persistency and relaxed (epoch) persistency. Whereas CPUs rely only on clwb (cache-line-write-back)/clflushopt (cache-line-flush-optimized) \cite{intel} instructions, we consider and compare GPU alternatives: store write-through ({\sf store.wt}) and {\sf l2wb} to flush all dirty blocks in the L2 cache. We found that a mixture of them is warranted: store.wt is profitable for writing undo logs without write allocation (as temporal locality is absent in failure-free execution) and for implementing strict persistency. We also found that l2wb is profitable in cases where it is difficult/expensive to re-generate addresses required for clwb. We also use the {\sf membar} instruction as a persist barrier between epochs. Since GPU memory controller WPQs are not in the non-volatile domain, we evaluate the effect of using pcommit instruction to flush blocks in WPQs to PM. 

Third, for the epoch persistency model, we make an important observation that the thread hierarchy in the GPU programming model provides intuitive choices for epoch granularity. Based on the characteristics of a kernel, we propose three epoch granularities: kernel level, CTA (cooperative thread array) level, and loop level. We show that epoch selection is crucial to performance, and different workloads require different epoch scopes for optimal tradeoff between performance and recoverability from crashes. 

Fourth, due to GPU application performance being bandwidth-sensitive and logging adds to bandwidth pressure, we need to reduce the reliance on the logging, both in frequency and the size of logs. We propose to leverage idempotency analysis \cite{igpu12, ido18} at different GPU thread hierarchy granularities and found that this analysis not only helps remove the need to create undo logging, but also reduces the size of logs when the epoch is not idempotent. We also show that many GPU kernels can benefit from this optimization.

Finally, we show evaluation results on two platforms, real GPUs (NVIDIA GTX1080) and simulated GPUs, that demonstrate data persistency and recoverability on GPUs can be achieved at low performance overheads. 

\section{Background and Related Work}
\label{sec:back}

\subsection{Memory Persistency Models on CPUs}
\label{sec:back_pm}

Byte-addressable NVM, aka persistent memory (PM), provides opportunities for high-performance in-memory recoverable data structures (RDS). However, due to write-back caches in CPUs, the durability order of writes to main memory can be different from the program order of stores. This does not present a problem for volatile memory such as DRAM. But it is not the case for PM. For example, assuming that  the data fields `$p \rightarrow data$' and `$p \rightarrow state$' reside in different cache lines and the update to `$p \rightarrow data$' precedes the update to `$p \rightarrow state$'. Due to the write-back last-level cache (LLC), `$p \rightarrow state$' in memory may be updated while `$p \rightarrow data$' has not. In this case, if a fault, e.g., a power failure, happens, the persistent memory state becomes incorrect after power is restored. To deal with this issue and support correct implementations of RDS, memory persistency models \cite{pelley15, pelley14} formally specify the order of writes to PM. 
In particular, strict persistency means that the persistent memory order is identical to the volatile memory order, which is governed by the memory consistency model. In comparison, relaxed persistency reduces the order constraints and two such persistency models, epoch and strand persistency, have been proposed \cite{pelley15, pelley14}. Under the epoch persistency model, the execution of a thread is separated into epochs separated by persist barriers.
The durability order of stores to different addresses is only enforced between epochs but not within an epoch. 
The order of conflicting accesses, i.e., accesses to the same address, on the other hand, is maintained, which is referred to as strong persistent atomicity. The strand persistency model relaxes the order constraints even further but requires programmers to express dependencies so as to remove unwanted memory ordering. Therefore, we do not consider strand persistency in this work.

The memory persistency models mentioned above specify the durability order of stores but ordering alone does not provide recoverability. For example, assume that with either strict or epoch persistency, `$p \rightarrow data$' and `$p \rightarrow state$' in PM are updated in the program order. But it is still possible that a fault happens after `$p \rightarrow data$' is updated but before `$p \rightarrow state$' being updated. In this scenario, the memory state in PM is still not correct for data recovery. What is missing here is transaction-like semantics, which requires that a group of stores are made durable together or none at all. To achieve such operations for PM, durable transactions~\cite{intel, kolli16, volos11} have been proposed. Through either redo or undo logging, they enable data to be recovered if a failure occurs during a transaction. The overhead of logging can be significant and there are some recent works to either reduce the size of the logs using recomputation~\cite{elnawawy17} or to improve the performance using hardware logging~\cite{atom17, shin17}.  

\subsection{GPU Architecture and Programming Model}
\label{sec:arch}


Modern GPUs employ the single-instruction multiple-thread (SIMT) architecture. 
A GPU consists of multiple streaming multiprocessors (SMs). On each SM, there is one or more warp schedulers to feed instructions to the ALU or memory pipelines. The GPU memory hierarchy includes register files, L1 D-caches, shared memory, constant caches, texture caches, and an unified L2 cache. The L2 cache contains multiple partitions and there is a memory controller for every one or two partitions so as to achieve high memory access bandwidth. 
The L1 D-caches typically do not use the write back (WB) policy (e.g., a write evict policy instead) and currently there is no coherence support among the L1 D-caches residing in multiple SMs. The L2 cache uses the WB policy while the GPU ISA may support the store instructions with an option to write through the L2 cache. 

The SIMT programming model is a single program multiple data (SPMD) model and massive threads are organized in a hierarchy. A kernel is launched with a grid of collaborative thread arrays (CTAs). A CTA in turn contains many threads, which can communicate and synchronize with each other through shared memory. Each thread/CTA determines its workload using its thread/CTA id. Each SM can host one or more CTAs depending on their resource usage. The threads in a CTA form warps, each of which is executed in the single-instruction multiple-data (SIMD) manner.

Memory consistency models have not been formally defined on GPUs \cite{singh15}. Until recently, Heterogeneous System Architecture (HSA) Foundation \cite{hsa15} and OpenCL \cite{opencl} start to adopt the C11's datarace-free-0 (DRF-0) model, which guarantees sequential consistency (SC) for data-race-free 
code, but is undefined for the cases with data-races. A few recent works show that the overhead of SC or TSO (Total Store Order) can be significantly reduced for GPUs~\cite{singh15, hlrc16, ren17}.

\section{GPU Memory Persistency}
\label{sec:gpu_pm}

We explore how to adapt and re-architect two memory persistency models for GPUs: strict and epoch persistency. We propose a compiler approach to facilitating programmers to utilize the persistency models. In the original source code, the programmer simply inserts {\sf pragmas} to annotate the desired persistency model along with the options of implementation. Different pragmas are used for different persistent models:
\begin{lstlisting}[numbers=none]
#pragma gpu_pm strict options
#pragma gpu_pm epoch epoch_scope options
\end{lstlisting}
Our compiler produces code for execution on real GPUs using existing GPU instructions, as well simulated GPUs with new instructions that we add. The compiler approach supports three scopes of an epoch, including kernel-level, CTA-level and loop-level epochs, so as to take advantage of the SIMT programming model. Epochs with different scopes from these three can also be realized with the same architectural support. For example, an user may explicitly specify a region of kernel code containing several loops or a region of host code containing multiple kernel invocations as an epoch using persistency barriers.

\subsection{Strict Persistency}
\label{sec:sp}

\begin{figure}
\begin{subfigure}[b]{\linewidth}
\begin{lstlisting}
lbm_kernel{
#pragma gpu_pm strict clwb
  ... = input[loc1(tid)];  ... = input[loc2(tid)];
  ... // compute...
  output[loc1(tid)] = ...; output[loc2(tid)] = ...;}
\end{lstlisting}
\squeezeupless
\caption{}
\squeezeupless
\end{subfigure}

\begin{subfigure}[b]{\linewidth}
\begin{lstlisting}
lbm_strict{
  ... = input[loc1(tid)];  ... = input[loc2(tid)];
  ... // compute...
  output[loc1(tid)] = ...;
  clwb(&output[loc1(tid)]); {sfence; pcommit;}  sfence;
  output[loc2(tid)] = ...;
  clwb(&output[loc2(tid)]); {sfence; pcommit;}  sfence;}
\end{lstlisting}
\squeezeupless
\caption{}
\end{subfigure}
\caption{An example for strict persistency. (a) Original code with pragma, (b) the code with the compiler added instructions for strict persistency.}
\squeezeup
\label{fig:code_sp}
\end{figure}

To support strict persistency for GPUs, we can persist the data in the program order. To do so, for each store, we add a clwb (or clflush/clfushopt) instruction to write the dirty cache line to the memory controller, a pcommit instruction to write the data in the WPQ to persistent memory if the WPQ in the memory controller is not durable, and sfence instructions to ensure the order of memory operations. One such example based on the benchmark lbm (see our methodology in Section \ref{sec:meth}) is shown in Figure \ref{fig:code_sp}. Figure \ref{fig:code_sp}(a) is the original code with a gpu\_pm pragma to direct the compiler to generate code for strict persistency. Figure \ref{fig:code_sp}(b) is the generated code. The instruction pair {sfence, pcommit} is not needed if the WPQs are durable.

Besides the clwb instruction, strict persistency can also be implemented using store.wt instructions. A user can choose this option by specifying `wt' in the pragma, in which case the compiler replaces all the store instructions with the store.wt instructions.

In current GPUs, {\sf membar/fence} instructions enforce memory ordering in NVIDIA PTX ISA~\cite{ptx}. The PTX manual states that ``{\em The membar instruction guarantees that prior memory accesses requested by this thread are performed at the specified level, before later memory operations requested by this thread following the membar instruction. The level qualifier specifies the set of threads that may observe the ordering effect of this operation.}" For the evaluation on real GPUs, we choose membar at the `gl' level, i.e., the GPU level, equivalent to the {\sf fence.gpu} instruction. For the evaluation on simulated GPUs, we implement the instruction with the semantics that all its prior memory operations from the same warp receive their acknowledgements. 

{\sf clwb} instruction does not exist in current GPUs, hence we use store.wt in its place for evaluation on real GPUs, and implement it on the simulated GPUs. There is no existing support for the pcommit instruction either in current GPUs. Therefore, we introduce this instruction. As there are multiple memory partitions and memory controllers, stores from the same warp with different addresses may be mapped to different memory controllers. As a result, in order to correctly implement the pcommit instruction, we need to drain the WPQs in all the memory controllers. We model such semantics in our simulator for the case when the WPQs are not durable. 

\subsection{Epoch Persistency}
\label{sec:ep}

As discussed in Section \ref{sec:arch}, the GPU/SIMT programming model requires programmers to explicitly specify the thread hierarchy. For HPC workloads, it is a common practice that each thread is used to compute one or few elements in the output domain and the threads in one CTA compute a tile/subblock of output elements. For example, in matrix multiplication, a thread is used to compute one or few elements in the product matrix. A CTA computes a tile of elements in the product matrix. For complex applications, the overall computation can be decoupled into multiple kernels. 

Epoch persistency requires choosing the scope of epochs. Since an epoch scope corresponds to the code region that needs to be re-executed on a failure, the scope of an epoch must correspond to a code region that the programmer finds easy to analyze and to reason about failure recovery. GPU thread hierarchy provides intuitive epoch granularities for this purpose: an entire kernel as an epoch, a CTA as an epoch, or a loop iteration as an epoch.  

To help users determine the proper epoch granularity, we propose the following scheme. First, based on its runtime characteristics, we classify a GPU kernel into one of the three categories: (1) short-running kernels; (2) long-running kernels with short-running CTAs; and (3) long-running kernels with long-running CTAs. Note that determining long-running vs. short-running needs to take the failure rate/mean-time-to-failure (MTTF) and recovery cost into consideration to ensure forward progress. Then we apply three different epoch persistency models accordingly: kernel-level epoch persistency for short-running kernels, CTA-level epoch persistency for long-running kernels with short-running CTAs, and loop-level epoch persistency for long-running kernels with long-running CTAs. 

\noindent\textbf{a. Kernel-Level Epoch Persistency}

\begin{figure}
\begin{subfigure}[b]{\linewidth}
\begin{lstlisting}
... //setup thread hierarchy, i.e., grid & block.
... // prepare input array
#pragma gpu_pm epoch kernel
histo_kernel_2<<<grid, block>>>(input, output);
... // consume output array
\end{lstlisting}
\squeezeupless
\caption{}
\squeezeupless
\end{subfigure}

\begin{subfigure}[b]{\linewidth}
\begin{lstlisting}
histo_kernel_2<<<grid, block>>>(input, output);
cudaDeviceSynchronize();
cudaL2WB(); 
cudaDeviceSynchronize(); //wait for l2wb to finish
... // consume output array
\end{lstlisting}
\squeezeupless
\caption{}
\end{subfigure}
\caption{An example for kernel-level epoch persistency.(a) Original code with pragma, (b) the code with compiler added APIs for kernel-level epoch persistency.}
\squeezeup
\label{fig:code_epk}
\end{figure}

For kernel-level epoch persistency, each kernel invocation is an epoch. At the end of the kernel execution, we persist all updated data and add a persist barrier. In current GPUs, the dirty data in the L2 cache are not written back to the device memory as the memory controller monitors the incoming data requests (e.g., from the host CPU) and feeds the most recent data from the L2 cache directly if needed. To support kernel-level epoch, we propose to add a new {\sf l2wb} instruction to write back {\em all} dirty lines in the L2 into device PM. This instruction can be used in either the kernel code or the host code through a driver API. Figure \ref{fig:code_epk} shows such an example based on the histogram benchmark. The original host code is shown in Figure \ref{fig:code_epk}(a). The pragma before the kernel launch is for the compiler to generate the host code with the added APIs, including the synchronization and l2wb, as shown in Figure \ref{fig:code_epk}(b). The device synchronization function, cudaDeviceSynchronize(), is used as the persist barrier between kernel invocations when there are multiple kernel invocations.

\noindent\textbf{b. CTA-Level Epoch Persistency}

\begin{algorithm}[t] 
\small {
\caption{Code generation for CTA-level epoch persistency with clwb}
\label{alg:compiler}
\begin{algorithmic}[1]
\Require{Kernel source code} 
\Function{EP-CTA-CLWB}{Kernel}
    \State{Create a post-dominant block in the end of kernel}
    \State{Move code generator to the created block}
    \State{Get all global memory stores in the kernel}
    \For{each store}                    
        \State{Detect use-define chain of the store address}
        \State{Replicate all statements in the chain if the address is no longer available}
        \State{Insert $clwb$ with the address}
    \EndFor
    \State{Insert $sfence$}
    \If{WPQ is volatile} 
        \State{Insert $pcommit$ and $sfence$}
    \EndIf
\EndFunction
\end{algorithmic}
}
\end{algorithm}


\begin{figure}
\begin{subfigure}[b]{\linewidth}
\begin{lstlisting}
lbm_kernel{
#pragma gpu_pm epoch cta clwb
  ... = input[loc1(tid)]; ... = input[loc2(tid)];
  ... // compute...
  output[loc1(tid)] = ...; output[loc2(tid)] = ...;}
\end{lstlisting}
\squeezeupless
\caption{}
\squeezeupless
\end{subfigure}

\begin{subfigure}[b]{\linewidth}
\begin{lstlisting}
lbm_CTA{
  ... = input[loc1(tid)];  ... = input[loc2(tid)];
  ... // compute
  output[loc1(tid)] = ...; output[loc2(tid)] = ...;
  clwb(&output[loc1(tid)]); clwb(&output[loc2(tid)]);
  {sfence; pcommit;}  sfence;}
\end{lstlisting}
\squeezeupless
\caption{}
\end{subfigure}
\squeezeup
\caption{An example for CTA-level epoch persistency.(a) Original code with pragma, (b) the code with the compiler added instructions for CTA-level epoch persistency using the clwb option.}
\squeezeup
\label{fig:code_epc}
\end{figure}

In the CTA-level epoch persistency, each CTA is an epoch. Durability ordering is not enforced for stores within a CTA and we just need to persist all the updated data at the end of each CTA. Many GPU applications, especially scientific computing workloads including BLAS, stencil, FFT, etc., share a popular programming pattern that the inputs are accessed at the beginning of a kernel function and the outputs are generated at the end, and many threads in a CTA collaboratively compute a set of output data. Such a programming pattern fits nicely with the CTA-level epoch persistency model. One such an example based on the benchmark lbm is shown in Figure \ref{fig:code_epc}. Figure \ref{fig:code_epc}(a) shows the original code with a pragma to indicate the compiler to generate the CTA-level epoch persistency code using the clwb option. And the code with the compiler-inserted instructions for persistency is shown in Figure \ref{fig:code_epc}(b). Compared to the code using the strict persistency model in Figure \ref{fig:code_sp}(b), the stores and the cache-line write backs are performed in an overlapped manner instead of being sequential. The last sfence instruction also serves as a persist barrier. 

The compiler algorithm for such code transformation is shown Algorithm \ref{alg:compiler}. It first creates a basic block that post-dominates all statements and this basic block is used for code generation. Then it determines all global memory stores in the kernel. For each store, the compiler detects the use-define chain of the store address. In the created basic block, the whole chain is replicated to re-calculate the address if necessary. And the clwb instruction is inserted with the address to write back the cache line. After all clwb instructions have been generated, the sfence instruction is inserted to wait for the clwb instructions to be posted.  

Besides the clwb option, we can also use wt and/or l2wb to implement CTA-level epoch persistency. In some kernel functions, their outputs are distributed in the code and it may be hard or too costly to re-generate the addresses at the end of the kernel, which are to be used by the clwb instructions. In such cases, we can either replace the store instructions with the stores using the WT operator (i.e., store.wt) or resort to the l2wb instruction followed by a sfence instruction to persist all the dirty cache lines in the L2 cache, even some of them are not updated by this particular CTA. 

Note that we argue that there is no need for a CTA-level synchronization, i.e., syncthreads(), after the last sfence instruction. The reason is that every warp in the CTA will execute the sfence instruction as its last instruction, which guarantees that no further memory instructions will be issued from this CTA.

With CTA-level epochs, there is one distinction between the epoch persistency model on CPU and GPU. As specified in the GPU programming model, CTAs are supposed to be executed in parallel without ordering constraints. As a result, we can view that with CTA-level epochs, there are multiple concurrent epochs running on a GPU. In constrast, in a sequential program, epochs are executed sequentially and there are order constraints between epochs. 

\begin{figure}
\begin{subfigure}[b]{\linewidth}
\begin{lstlisting}
tpacf_kernel(g_hists, data) {
  __shared__ int s_hists[N_BINS][N_THD];
  ... // Initialization
  // Long nested loop
#pragma gpu_pm epoch loop l2wb
  for (i = 0; i < N_ELEMS; i += CTA_SIZE) {
    for (k = 0; k < CTA_SIZE; k++) {
      ...
      bin_idx = ...
      s_hists[bin_idx][tid] += 1; 
    } ... }
\end{lstlisting}
\squeezeupless
\caption{}
\squeezeupless
\end{subfigure}

\begin{subfigure}[b]{\linewidth}
\begin{lstlisting}
__device__ int shadow[N_CTA][N_BINS][N_THD]
tpacf_kernel_loop(g_hists, data) {
  __shared__ int s_hists[N_BINS][N_THD];
  ... // Initialization
  // Long nested loop
  for (i = 0; i < N_ELEMS; i += CTA_SIZE) {
    for (k = 0; k < CTA_SIZE; k++) {
      ...
      bin_idx = ...
      s_hists[bin_idx][tid] += 1; 
      shadow[cta_id][bin_idx][tid] = s_hist[bin_idx][tid];
      //The next two lines are for strict persistency
      //clwb(&shadow[cta_id][bin_idx][tid]); |\label{line:sp1}|
      //{sfence; pcommit;} sfence;           |\label{line:sp2}|
    } //end of the inner loop 
    ... 
    if( tid == 0) l2wb; // write back L2 dirty cache lines 
    {sfence; pcommit;} sfence;
  } //end of the outer loop
...}
\end{lstlisting}
\squeezeupless
\caption{}
\end{subfigure}
\squeezeup
\caption{An example for loop-level epoch persistency. The scratchpad memory is made persistent through a shadow copy in global memory. (a) Original code with pragma, (b) the code with the compiler added instructions for loop-level epoch persistency using the l2wb option.}
\squeezeup
\label{fig:code_epl}
\end{figure}

\noindent\textbf{c. Loop-Level Epoch Persistency}

In the loop-level epoch persistency model, the scope of an epoch is reduced to an iteration of a long-running loop in a kernel. Here, we use the benchmark tpacf as a case study. The simplified kernel code is shown in Figure \ref{fig:code_epl}(a) and contains a long-running nested loop. A pragma is inserted immediately before the outer loop to indicate the scope of the epoch. Also, because the code uses shared memory for the intermediate results in every loop iteration. The compiler creates a shadow copy of the shared memory array for each CTA and persists the shadow copy for shared memory data recovery. As the shared memory array is updated in every iteration in the innermost loop, it is very costly to reconstruct such array indices at the end of the outermost loop. Therefore, we can leverage the l2wb instruction to write back all the L2 dirty lines, which is then followed by a sfence instruction as the persist barrier to ensure that no subsequent memory operations can be issued from this warp until the write backs are finished. Note that due to the costly overhead of the l2wb instruction,
we only use one thread in a CTA to execute this instruction. The resulting code is shown in Figure \ref{fig:code_epl}(b). For reference, we also include the (commented out) code for strict persistency in lines \ref{line:sp1} \& \ref{line:sp2}. 

With loop-level epoch persistency, the order constraints are between the epochs (i.e., loop iterations) in a single warp while epochs in different warps and CTAs can be executed in parallel.

\noindent\textbf{d. Summary}

\begin{table}[]
\caption{A summary of memory persistency models, the architectural support, and the targeted kernels.}
\centering
\scalebox{0.96} {
\small
\begin{tabular}{|c|c|c|c|c|}
\hline
\multirow{2}{*}{\begin{tabular}[c]{@{}c@{}}Persistency\\ Models\end{tabular}} & \multirow{2}{*}{\begin{tabular}[c]{@{}c@{}}Strict\\ Persistency\end{tabular}}         & \multicolumn{3}{c|}{\begin{tabular}[c]{@{}c@{}}Relaxed Persistency\end{tabular}}                                                                                                                                                                                      \\ \cline{3-5} 
                                                                              &                                                                                       & \begin{tabular}[c]{@{}c@{}}Kernel-Level Epoch\end{tabular}            & \begin{tabular}[c]{@{}c@{}}CTA-Level Epoch\end{tabular}                                     & \begin{tabular}[c]{@{}c@{}}Loop-Level Epoch\end{tabular}                                    \\ \hline
\begin{tabular}[c]{@{}c@{}}Architectural\\  Support\end{tabular}              & \begin{tabular}[c]{@{}c@{}}clwb/clflush(opt);\\store.wt;  sfence; pcommit\end{tabular} & \begin{tabular}[c]{@{}c@{}}l2wb;\\   DeviceSynchronization\end{tabular} & \begin{tabular}[c]{@{}c@{}}clwb/clflush(opt); store.wt; \\ sfence; pcommit; l2wb\end{tabular} & \begin{tabular}[c]{@{}c@{}}clwb/clflush(opt); store.wt; \\ sfence; pcommit; l2wb\end{tabular} \\ \hline
Suitable Kernel                                                               & All                                                                                   & Short-running kernels                                                   & \begin{tabular}[c]{@{}c@{}}Long-running kernels \\ with short-running CTAs\end{tabular}       & \begin{tabular}[c]{@{}c@{}}Long-running kernels \\ with long-running CTAs\end{tabular}        \\ \hline
\end{tabular}
}
\squeezeup
\label{tab:models}
\end{table}

We explore both strict and epoch persistency models for GPUs. A summary of their architectural support and their targeted GPU kernels is presented in Table \ref{tab:models}. 
Note that different models treat shared memory data (i.e., the data in the scratchpad memory) differently. Among the epoch persistency models, only the ones with the scope less than a CTA, e.g., the loop-level, need to construct a shadow copy in the global memory and persist the data at the end of an epoch. For the CTA- and kernel-level, the shared memory data are no longer live at the end of an epoch, therefore there is no need for the data to be persisted.

\section{Durable Transactions for GPUs}
\label{sec:gpu_dt}

The memory persistency models (Section \ref{sec:gpu_pm}) specify the durability ordering of stores, which is necessary but insufficient for guaranteeing a fail-safe state as discussed in Section \ref{sec:back_pm}. 
Durable transactions are often required to persist a group of stores together or none at all. In this section, we discuss turning an epoch into a durable transaction with undo logging, or in some cases omit logging altogether by exploiting the idempotency property of an epoch. 

Software-based undo logging contains the following steps. First, before a transaction starts, an undo log is created by making a copy of the data to be updated and this undo log is persisted. Second, we set and persist a flag to indicate that the transaction is running. Third, during the transaction, data are updated and at the end of the transaction, the updated data are persisted. Fourth, we mark the transaction complete and release the undo log.

With undo logging, the recovery code checks the flags to see whether there is a transaction is interrupted. If so, it uses the undo log to restore the data.

\subsection{Kernel-Level Durable Transactions}
\label{sec:dtk}

\begin{figure}
\begin{lstlisting}
enum flag {initial, inTx, Complete};
... //setup thread hierarchy, i.e., grid & block.
... // prepare input array
cudaMemcpy(undo_log, output,  
         size, cudaMemcpyDeviceToDevice); // create undo log
flag = inTx; 
clwb(&flag); sfence;  //persist the flag in host memory
cudaDeviceSynchronize();  // wait for cudaMemcpy to finish
histo_kernel_2<<<grid, block>>>(input, output);
cudaDeviceSynchronize(); // wait for the kernel to finish

cudaL2WB();
cudaDeviceSynchronize(); //wait for l2wb to finish
flag = Complete; 
clwb(&flag); sfence; //persist the flag in host memory
... // consume output array
\end{lstlisting}
\caption{A code example for kernel-level durable transaction.}
\squeezeup
\label{fig:code_dtk}
\end{figure}

As GPUs are used as accelerators, their input data are prepared at the host side and copied to the device. Then, the kernel is invoked by the host. Therefore, we propose to implement kernel-level durable transactions in the host code. We also assume that the host side memory is persistent. The resulting code based on the histogram benchmark is shown in Figure \ref{fig:code_dtk}. We define a flag to show whether a transaction is running (`inTx') or completed. A copy of the data to be updated (i.e., output) is persisted in the device memory using the `cudaMemCpy' function. Then, the flag is set to be inside a transaction (`inTx') and persisted in host PM. After the undo log is persisted, the kernel is launched. Next, after the kernel completes and persists its results using the l2wb instruction, the flag is set to `complete' and persisted in host PM. With this transaction-style execution, the recovery code at the host side checks the flag. If it is `inTx', the potentially corrupted data (i.e., output) is restored using the undo log.

The kernel `histo\_kernel\_2' in Figure \ref{fig:code_dtk} computes a histogram of the input and does not change the input. Also, there are no side effects during kernel execution. Therefore, the kernel is \textit{idempotent}, meaning that it can be executed multiple time without changing the result. We propose to leverage re-execution to recover from failure rather than using the undo log. As a result, we can completely eliminate the code for undo logging (i.e., `cudaMemCpy' and the first `cudaDeviceSyncronize') in Figure \ref{fig:code_dtk}. 

In some kernel functions, the input and the output may be altered. In this case, the undo log needs to include the input as well. If we use the non-unified memory model, i.e., the host code explicitly copies the data to the device memory, a copy of the input should already exist in host memory. Therefore, we do not need to make a redundant copy of the input data in device memory. On the other hand, if the unified memory model is used, we need to explicitly make a copy of the input, either in host or device PM.

\subsection{CTA-Level Durable Transactions}
\label{sec:dtc}

With a CTA as a transaction, undo logging is implemented at the device 
side. Using the benchmark lbm as an example, the kernel function with undo logging at the CTA level is shown in Figure \ref{fig:code_dtc}(a). 
We first create an undo log for the output elements that are to be updated by the CTA. Here, we use the store.wt option as an alternative to the regular store instruction followed by  clwb since the log has no reuse in failure-free execution. After ensuring that all the threads persist their log using the sfence followed by the `syncthreads()' function, we set the flag to be `inTx' and make it durable. Then at the end of the CTA, we ensure that all the outputs have been persisted using another syncthreads() and set the flag to be `complete'.

\begin{figure}
\begin{subfigure}[b]{\linewidth}
\tiny{
\begin{lstlisting}
enum FLAG {initial, inTx, complete}; FLAG flag[NUM_CTA];
lbm_CTA_log{
  ... = input[loc1(tid)];  ... = input[loc2(tid)];
  //log[loc1[tid]] = output[loc1[tid]]; clwb(&log[loc1[tid]]);
  st.wt &log[loc1(tid)], output[loc1(tid)];   
  st.wt &log[loc2(tid)], output[loc2(tid)]; 
  // compute...
  {sfence; pcommit;} sfence; 
  __syncthreads(); // logs are durable
  if (tid == 0) {
    st.wt &flag[ctaid], inTx; // inside tx
    {sfence; pcommit;} sfence;  }
  output[loc1(tid)] = ...; output[loc2(tid)] = ...;
  clwb(&output[loc1(tid)]); clwb(&output[loc2(tid)]);
  {sfence; pcommit;} sfence; 
  __syncthreads(); // CTA is done
  if (tid == 0) {
    st.wt &flag[ctaid], complete; // committed
    {sfence; pcommit;} sfence;}}
\end{lstlisting}
}
\squeezeupless
\caption{}
\squeezeupless
\end{subfigure}

\begin{subfigure}[b]{\linewidth}
\tiny{
\begin{lstlisting}
lbm_CTA_idem{
  ... = input[loc1(tid)]; ... = input[loc2(tid)];
  st.wt &flag[ctaid], inTx; // inside tx
  {sfence; pcommit;} sfence;  
  // compute...
  output[loc1(tid)] = ...; output[loc2(tid)] = ...;
  clwb(&output[loc1(tid)]); clwb(&output[loc2(tid)]);
  {sfence; pcommit;} sfence; 
  __syncthreads(); // CTA is done
  if (tid == 0) {
    st.wt &flag[ctaid], complete; // committed
    {sfence; pcommit;} sfence;  }}
\end{lstlisting}
}
\squeezeupless
\caption{}
\end{subfigure}
\squeezeup
\caption{An example for CTA-level durable transaction. (a) The code with undo logging, (b) the optimized code using idempotency analysis.}
\squeezeup
\label{fig:code_dtc}
\end{figure}


When a CTA is idempotent, i.e., the kernel function has no anti data dependency and is re-executable, the undo log can be eliminated.~\footnote{Even when a CTA is not idempotent, idempotency analysis shows which stores can be safely repeated, hence we still apply it in order to reduce the size of undo logs.} We only need to set the flag of each CTA as shown in Figure \ref{fig:code_dtc}(b). In this case, the recovery code simply re-executes those CTAs with their flag being `inTx' and does not need to undo the changes using the undo log. Note that in Figure \ref{fig:code_dtc}(b), when we set the flag to `inTx' (i.e., 1), all the threads in a CTA will execute the code instead of using only thread 0 as in Figure \ref{fig:code_dtc}(a). The reason is that there is no syncthreads() function right before it. As a result, if only thread 0 updates this flag, there may be a chance that some threads/warps in the CTA change the output before thread 0 sets the flag, violating the transaction semantics. By allowing all the threads to set the flag, it ensures that the flag is set before any thread can change the output. Since all the threads set the same value to the flag, this data race is benign. 


\subsection{Loop-Level Durable Transactions}
\label{sec:dtl}

\begin{figure}
\begin{lstlisting}
__device__ int log[N_CTA][N_BINS][N_HISTS];
__device__ int flag[N_CTA];
__device__ int last_iter[CTA]; // last persisted iteration
__device__ int last_log_iter[N_CTA]; // last logged iteration
tpacf_kernel_loop_log(g_hists, data) {
  __shared__ int s_hists[N_BINS][N_THD];
  ... // Initiation
  // Long nested loop
  for (i = 0; i < N_ELEMS; i += CTA_SIZE) {
    // Create the log 
    for (b = 0; b < N_BINS; b++) 
      st.wt &log[cta_id][b][tid], shadow[cta_id][b][tid];
    st.wt &last_log_iter[cta_id], last_iter[cta_id];
    {sfence; pcommit;} sfence; 
    __syncthreads(); // log is durable for the CTA
    if (tid == 0) 
      st.wt &flag[cta_id], inTx; // inside Tx
    {sfence; pcommit;} sfence; 
    for (k = 0; k < CTA_SIZE; k++) {
      bin_idx = calculate(data[k]);
      s_hists[bin_idx][tid] += 1;
      shadow[cta_id][bin_idx][tid] = s_hist[bin_idx][tid];
    } //end of the inner loop
    ...
    if(tid == 0) l2wb; 
    st.wt &last_iter[cta_id], i; 
    {sfence; pcommit;} sfence;  
    __syncthreads(); // the results are durable
    if (tid == 0) st.wt &flag[cta_id], complete;
    {sfence; pcommit;} sfence; // committed
  } //end of the outer loop
  ...}

\end{lstlisting}
\caption{A code example for loop-level durable transaction.}
\squeezeup
\label{fig:code_dtl}
\end{figure}

To achieve loop-level durable transactions, we need to analyze the long-running loops in a kernel. The simplified kernel code of the benchmark tpacf is used a case study. The long-running outer loop uses scratchpad memory for data communication among threads within a CTA and is not idempotent. As discussed in Section \ref{sec:ep}c, we use a shadow copy of shared memory variables in global memory and this shadow copy is updated in each iteration. Therefore, we need to create an undo log for this shadow copy. Moreover, besides the flag, we need to record the meta data such as the loop iterator value to indicate which iteration is being executed. The resulting code is shown in Figure \ref{fig:code_dtl}. As the loop is not idempotent, the log cannot be eliminated and its overhead due to the increased memory traffic can be significant.


In comparison, for the kernels without data communication among threads in a CTA, i.e., each thread works on its private data, loop-level undo logging is relatively simple. 
Each thread backs up \& persists its private data to be changed at the beginning of the loop iteration and sets the flag to be `inTx'. Then, at the end of the loop iteration, the updated data are persisted and the flag is set to be `complete'. As there is no shared data, there is also no need for the syncthreads() barrier.

\section{Experimental Methodology}
\label{sec:meth}

We evaluate our proposed schemes on both an NVIDIA GTX1080 GPU and the GPGPU-Sim \cite{sim09}, a cycle-accurate GPU microarchitecture simulator. The GTX1080 GPU is hosted on a Red Hat 7.4 Linux machine and we use the CUDA 9.0 in our experiments. The simulation configurations of GPGPU-Sim 
are shown in Table \ref{tab:config}. 

Our experiments use all the benchmarks in the Parboil GPU benchmark suite \cite{parb12}. The kernels are listed in Table \ref{tab:bench}. As each benchmark may have multiple kernels, a number followed by a benchmark name is used to denote the order of the kernel in the benchmark. For each kernel, we also report whether the kernel function is idempotent in Table \ref{tab:bench}. We classify all the kernels into one of the three categories according to their execution time. If a kernel's execution time is less than 100us, it is categorized as a short-running kernel and is labelled `S'. The long-running kernels with short-running CTAs are labelled `LS' and they have the kernel execution time longer than 100us while the average execution time of their CTAs is less than 100us. When a kernel has CTAs that have an average execution time longer than 100us, it is categorized as a long-running kernel with long running CTAs or `LL'. Note that this classification is ad hoc and should take MTTF and the recovery cost into consideration. We use this setting for two reasons. The first is that it enables us to examine the performance impacts of different persistency models on a variety of kernels and evaluate the effects of our proposed optimizations. With a more realistic setting (e.g., in the order of seconds or minutes), all the kernels would be classified as short-running ones. Second, the 100us threshold used in our classification criteria implies an unreliable system as we need to achieve durable transactions with similar latency. Considering such an unreliable system with volatile memory, in order to make forward progress, we may resort to periodical checkpointing and recovery. A checkpoint would consist of the GPU context and the memory content, and would need to be persisted in host memory to ensure reliability, for which the latency would be much higher than 100us considering the PCIe bandwidth and the cost of GPU context switching. In other words, while volatile memory would not be able to support such fine-grain checkpoints, a GPU with PM along our proposed architectural support can achieve such level of durable transactions with relatively low performance overhead. With a coarser-grain epoch/durable transaction, the performance overhead would be further reduced.


\begin{table}[]
\caption{Baseline architecture configuration.}
\centering
\small
\begin{tabular}{|p{4cm}|p{10cm}|}
\hline
\# of SMs              & 20, SIMD width=32, 1.8GHz                                                    \\ \hline
Per-SM warp schedulers & 4 Greedy-Then-Oldest schedulers                                              \\ \hline
Per-SM limit           & 2048 threads, 64 warps, 32 CTAs                                   \\ \hline
Per-SM L1D-cache       & 24KB, 128B line, 6-way associativity, 256 MSHRs                                         \\ \hline
Per-SM SMEM            & 96KB, 32 banks                                                               \\ \hline
Unified L2 cache       & 2048 KB, 128KB/partition, 128B line, 16-way associativity, 256 MSHRs         \\ \hline
L1D/L2 policies        & xor-indexing, allocate-on-miss, LRU, L1D:WEWN,  L2: WBWA                     \\ \hline
Interconnect           & 16*16 crossbar, 32B flit size, 1.4GHz                                        \\ \hline
Memory Controller      & 8 channels, 2 L2 banks/channel, FR-FCFS scheduler, 1.2GHz, BW: 307GB/s \\ \hline
NVMM latency           & Read: 160ns, write: 480ns                                                    \\ \hline
DRAM Latency           & Read: 160ns, write: 160 ns                                                   \\ \hline
\end{tabular}
\label{tab:config}
\end{table}

\begin{table}[]
\caption{Benchmarks}
\centering
\small
\begin{tabular}{|c|c|c||c|c|c|}
\hline
Kernel & Type & Idempotent & Kernel  & Type & Idempotent \\ \hline \hline
bfs-1  & S    & No         & sad-1   & LS   & Yes        \\ \hline
bfs-2  & LL   & No         & sad-2   & S    & Yes        \\ \hline
cutcp  & LS   & Yes        & stencil & LS   & Yes        \\ \hline
grid-1 & LS   & No         & tpacf   & LL   & No         \\ \hline
grid-2 & LS   & Yes        & histo-1 & S    & No         \\ \hline
grid-3 & LS   & No         & histo-2 & S    & Yes        \\ \hline
grid-4 & LS   & Yes        & histo-3 & LS   & No         \\ \hline
grid-5 & S    & No         & histo-4 & S    & Yes        \\ \hline
gird-6 & S    & No         & lbm     & LS   & Yes        \\ \hline
mriq-1 & S    & Yes        & spmv    & LS   & Yes        \\ \hline
mriq-2 & LS   & No         & sgemm   & LS   & No         \\ \hline 
\end{tabular}
\label{tab:bench}
\end{table}

In our implementation, our compiler uses inline assembly to insert the new instructions including clwb, pcommit, and l2wb, into the kernel code. We also modify GPGPUsim to support the semantics of these instructions. The sfence instruction is implemented using the membar instruction. In modeling the clwb or l2wb instruction in GPGPUsim, the instruction is sent through the interconnect network to the L2 cache. The l2wb instruction is implemented with the controllers in multiple partitions, which go through every cache line and write back the dirty ones. It blocks subsequent L2 accesses until all the cache lines are checked, resulting in high performance overhead. The clwb instruction writes the specific dirty line in the L2 cache to a WPQ. If the target line is not dirty, no action will be taken except sending back an acknowledgement. If the target is dirty, an acknowledgement is sent back from a memory controller when the dirty cache line reaches the WPQ. The pcommit instruction needs acknowledgements from all the memory controllers when their WPQs are completely drained. For store.wt, our simulator models that its acknowledgment is sent by the memory controller once the data is written to device memory. Therefore, store.wt has higher latency than a clwb instruction in our simulation and uses volatile WPQs by default.

On the GTX1080 GPU, we use store.wt to enforce the store data to be written to memory and the membar.gl instruction as the sfence instruction. As the instructions l2wb and pcommit are not supported on the real GPU, we do not include them in the benchmark code, i.e., ignoring their overheads. Also, as DRAM is used as device memory, the read and write speeds do not reflect the characteristics of NVM. Nevertheless, we run our benchmarks on real GPUs to verify the functional correctness of our compiler generated code and compare the performance trend with our simulation results.

\section{Experimental Results}
\label{sec:eval}

\begin{figure}[t]
\centering
\includegraphics[width=.6\linewidth, trim=2 2 2 2, clip]{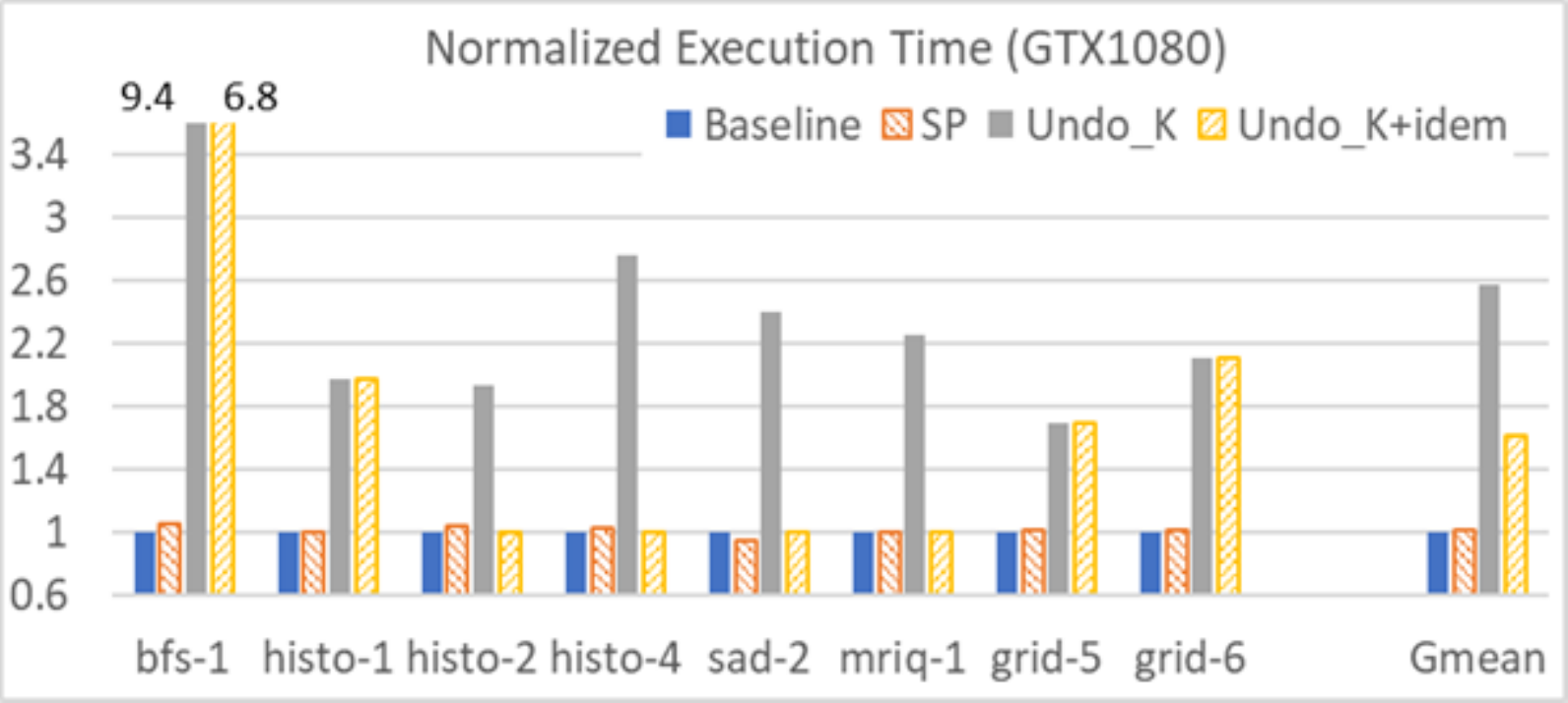}
\caption{Normalized execution time of short-running kernels with different persistency models on a GTX 1080 GPU (the lower, the better).}
\squeezeup
\label{fig:eval_sk_gtx}
\end{figure}

\begin{figure}[t]
\centering
\includegraphics[width=.6\linewidth, trim=2 2 2 2, clip]{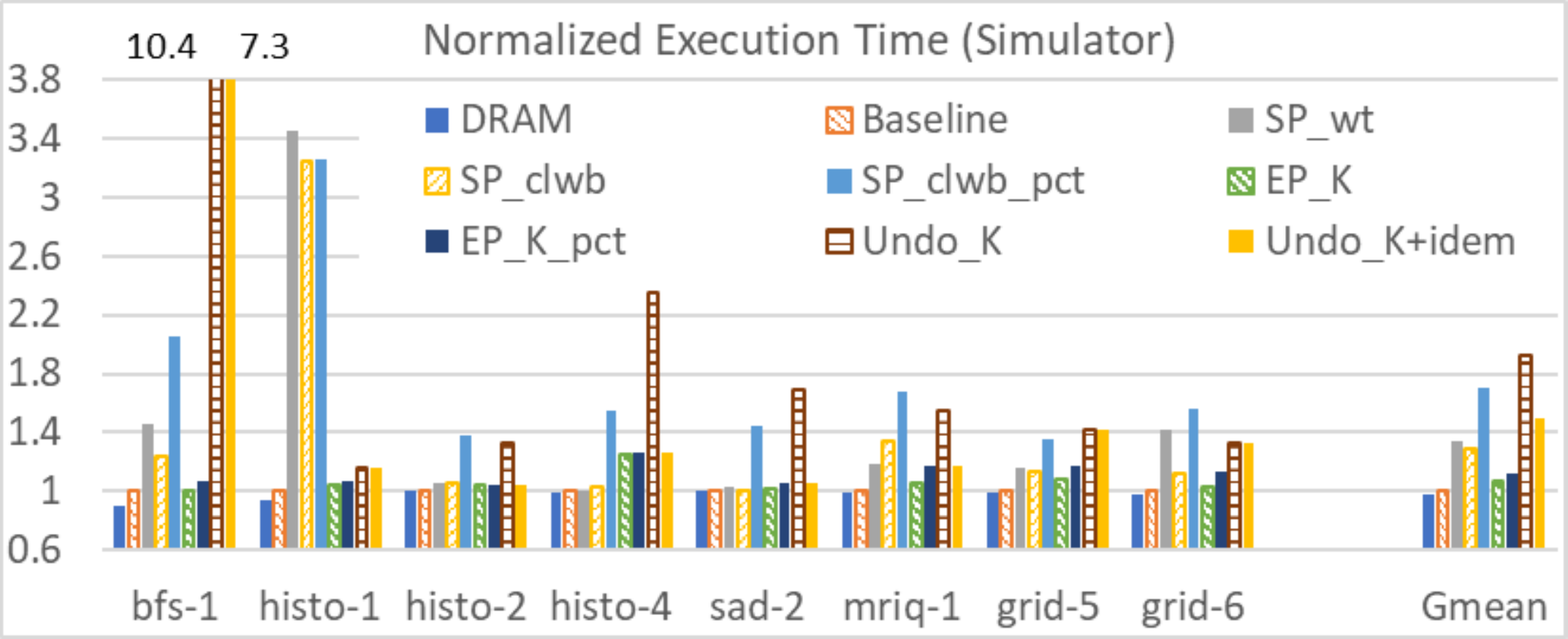}
\caption{Normalized execution time for short-running kernels with different persistency models on GPGPUsim.}
\label{fig:eval_sk_sim}
\squeezeup
\end{figure}

\begin{figure}[t]
\centering
\includegraphics[width=.6\linewidth, trim=2 2 2 2, clip]{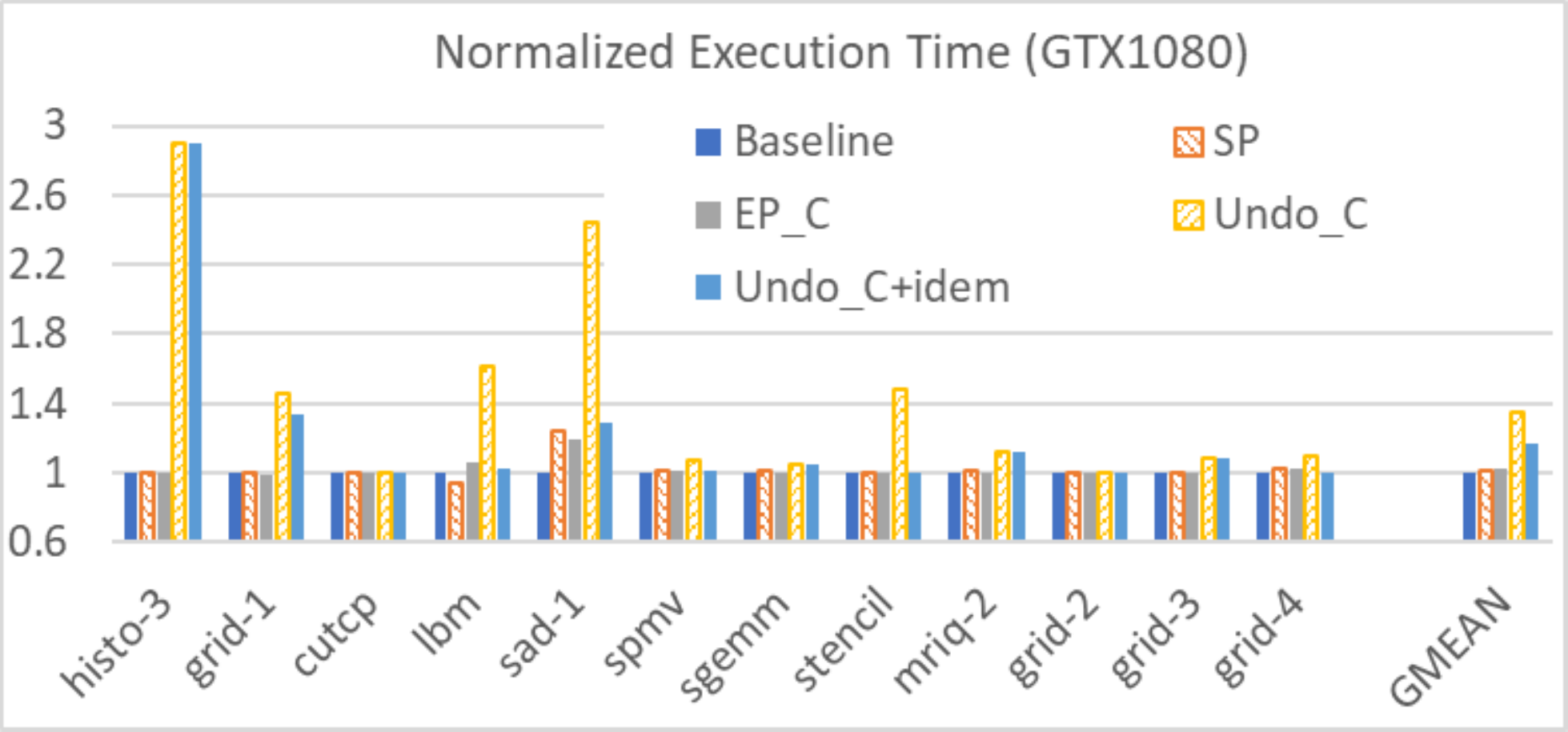}
\caption{Normalized execution time of long-running kernels with short-running CTAs using different persistency models on a GTX1080 GPU.}
\label{fig:eval_lsk_gtx}
\squeezeup
\end{figure}

\begin{figure*}[t]
\centering
\includegraphics[width=\linewidth, trim=2 2 2 2, clip]{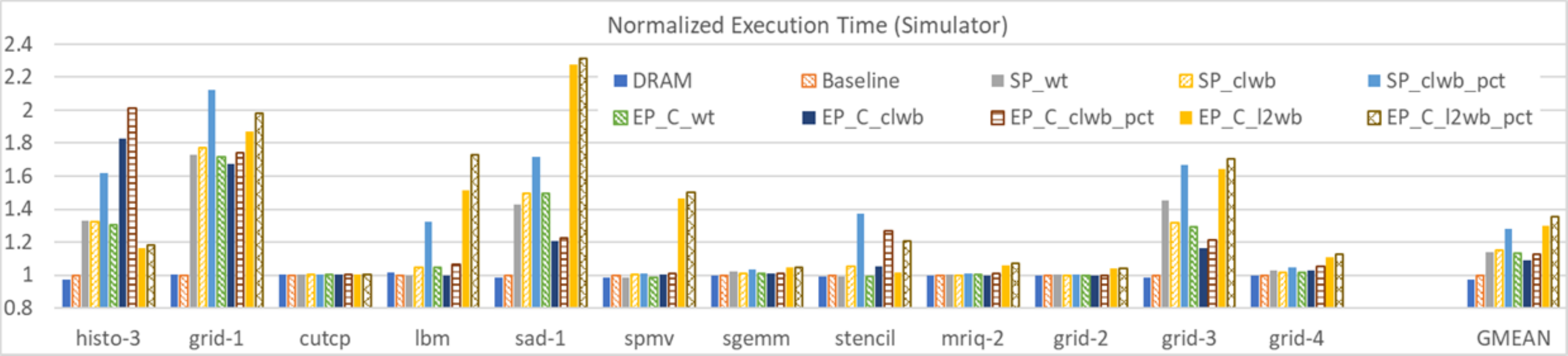}
\caption{Normalized execution time of long-running kernels with short-running CTAs using various persistency models on GPGPUsim.}
\label{fig:eval_lsk_sim}
\squeezeup
\end{figure*}


In our evaluation, we use the following the naming convention. We use `Baseline'/`DRAM' to denote the baseline using NVM/DRAM without persistency support. Among the persistency models, `SP' denotes strict persistency while `EP\_scope' denotes epoch persistency with a particular scope, which can be `K' (kernel level), `C' (CTA-level), or `L' (loop level). Among the durable transaction models, `Undo\_scope' denotes the undo logging with a particular scope, which adopts the same scope notation as in epoch persistency. We use `idem' to indicate that idempotency analysis is used to optimize undo logging. On the real GPU, store.wt is the only option to persist the memory stores. In the simulation results, we include the `wt', `clwb' and `l2wb' options to denote that the store.wt, clwb and l2wb instructions are used to persist the data, respectively. The label `pct' is included when pcommit instructions are used for volatile WPQs. 

\subsection{
Short-Running Kernels}
\label{sec:eval_sk}

\begin{figure}
\centering
\includegraphics[width=.6\linewidth, trim=2 2 2 2, clip]{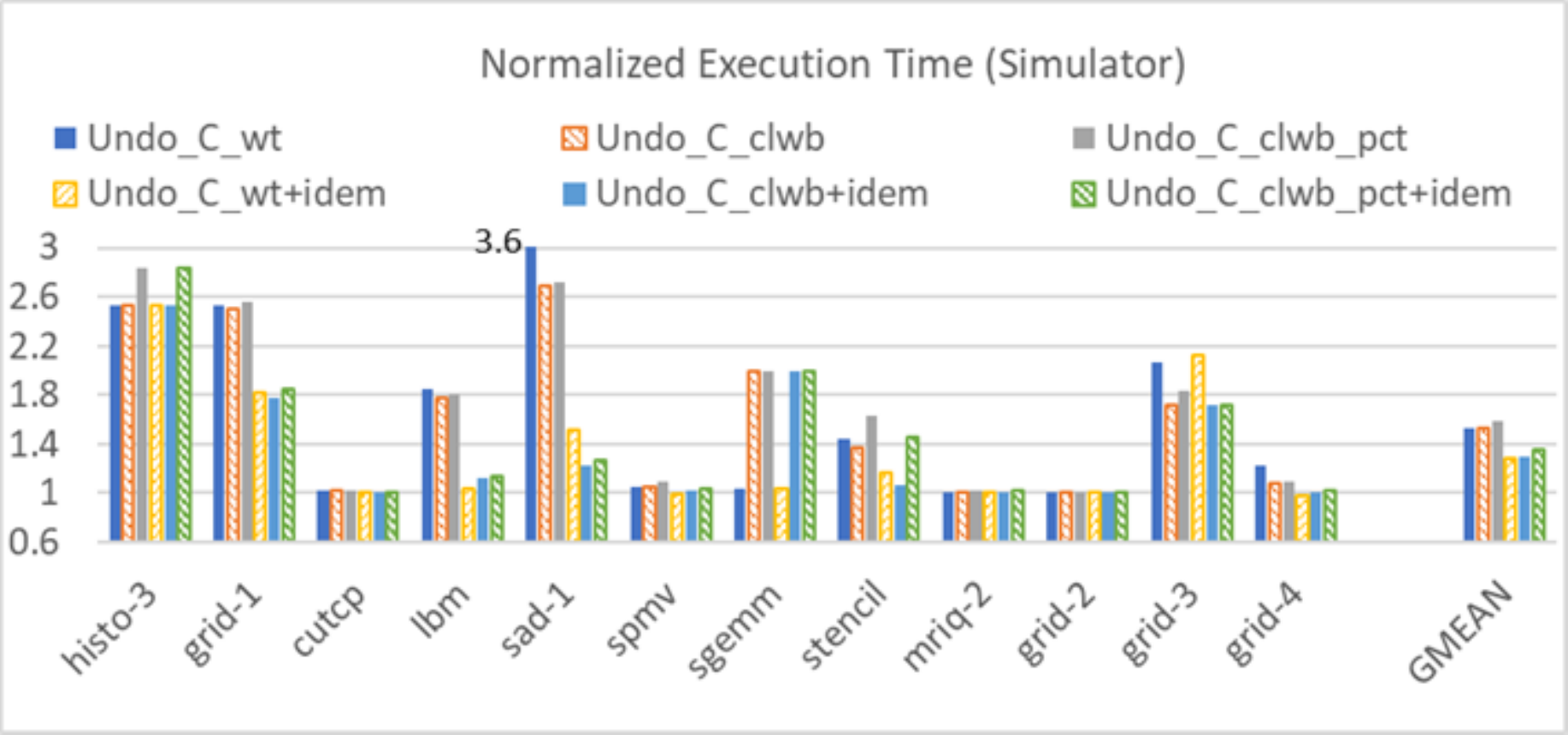}
\caption{Normalized execution time of long-running kernels with short-running CTAs using various durable transaction models on GPGPUsim.}
\squeezeup
\label{fig:eval_lsk_sim2}
\end{figure}

We first report the performance results of the short-running kernels on the GTX1080 GPU in Figure \ref{fig:eval_sk_gtx}. For each kernel, we show the normalized execution time to the baseline. The performance of kernel-level epoch persistency is the same as the baseline as we cannot include the overhead of l2wb on the real GPU. Several observations can be made from the figure. First, the performance impact of the WT operator and the membar instruction is rather limited, 1.5\% on average. Second, undo logging has high overhead although we use the high bandwidth device memory rather than host-side system memory. Third, idempotency analysis eliminates the undo logging overhead if a kernel is idempotent. Fourth, even if a kernel is not completely idempotent, idempotency analysis may still reduce the logging overhead. For example, the kernel bfs-1 shows the high logging overhead, which is due to its short kernel execution time compared to the memory copy time (which in turn indicates that the scope of this kernel is too small as an epoch). Although the kernel is not idempotent, it does not change all its inputs. Idempotency analysis discovers the opportunity and reduces the size of the undo log, thereby reducing the overhead. 

The performance results of the short-running kernels on the simulator are shown in Figure \ref{fig:eval_sk_sim}. 
The following observations can be made. First, the results correlate well those obtained from the real GPU, thereby confirming the observations made from Figure \ref{fig:eval_sk_gtx}. Second, there is very little performance difference (2.8\% on average) between the baseline with DRAM and the baseline with NVM, meaning that the additional write latency has small performance impact in the baseline. Third, when clwb instructions are used to send cache lines to WPQs, durable WPQs show significant performance improvement on average due to the reduced latency and the removal of pcommit+sfence instructions. Fourth, SP\_wt has close performance to SP\_clwb, meaning that SP\_wt achieves good performance without durable WPQs. The reason is that the clwb instructions introduce additional traffic through the interconnect network whereas the wt operator is part of the store instruction and does not incur additional traffic. Fifth, the kernel-level epoch persistency models have high parallelism due to the l2wb instruction. Therefore, the overhead of the kernel-level epoch persistency model over the baseline is low, 6.2\% and 11.6\% on average for EP\_K (epoch model with durable WPQs) and EP\_K\_pct (with volatile WPQs), respectively.
\subsection{
Long-Running Kernels with Short-Running CTAs}
\label{sec:eval_lsk}

We first report the performance of long-running kernels with short-running CTAs on the GTX 1080 GPU in Figure \ref{fig:eval_lsk_gtx}. 
Among the kernels, histo-3 and grid-1 make use of atomic operations on global memory variables. As a result, the CTA-level durable transaction model is not feasible for these two kernels. 
Therefore, we resort to kernel-level durable transactions for them. From the figure, we can observe: (a) minor overhead of the strict persistency and CTA-level epoch persistency models, (b) relatively high overhead due to undo logging, and (c) significant reduction in the undo-log sizes and performance overhead (from 35.2\% to 17.0\%) through idempotency analysis.

The simulation results of the persistency models and durable transaction models are shown in Figure \ref{fig:eval_lsk_sim} and Figure \ref{fig:eval_lsk_sim2}, respectively. 
Compared to the GTX1080 results, Figure \ref{fig:eval_lsk_sim} and \ref{fig:eval_lsk_sim2} confirms that strict persistency can be supported with relatively low overhead using either the in-place store.wt with volatile WPQs or clwb with durable WPQs. Also, the idempotency analysis effectively reduces the performance overhead of logging as observed on both the real GPUs and the simulator.

Figure \ref{fig:eval_lsk_sim} also shows that the CTA-level epoch persistency models have lower performance overhead than strict persistency models, especially when clwb instructions are used with volatile WPQs. The reason is that overlapping multiple memory writes as in the CTA-level epoch models saves more clock cycles than the sequential updates as in the strict persistency models. Due to such overlapping, the performance impact of the durable WPQs is also limited in the CTA-level epoch persistency models (i.e., EP\_C\_clwb vs. EP\_C\_clwb\_pct).

Between the epoch models, EP\_C\_clwb\_pct and EP\_C\_wt, some kernels show interesting behavior although both models use volatile WPQs. The kernel sad-1 shows better performance with EP\_C\_clwb\_pct while the kernel stencil shows better performance with EP\_C\_wt. The reason is that sad-1 has streaming-like memory updates but has poor memory coalescing. As a result, write-back caches can leverage spatial locality to reduce the number of memory updates, thereby achieving better performance using EP\_C\_clwb\_pct. On the other hand, the stencil kernel only has one store at the end of the kernel. Therefore, the CTA-level epoch persistency model is the same as the strict persistency model. As the store always misses the L2 cache, the write-not-allocate policy used with store.wt has lower latency than the write-back write-allocate policy. The overhead of the clwb \& pcommit instructions also contributes to the lower performance in EP\_C\_clwb\_pct than EP\_C\_wt.  

\subsection{
Long-Running Kernels with Long-Running CTAs}
\label{sec:eval_llk}

\begin{figure}
\centering
\includegraphics[width=.6\linewidth, trim=2 2 2 2, clip]{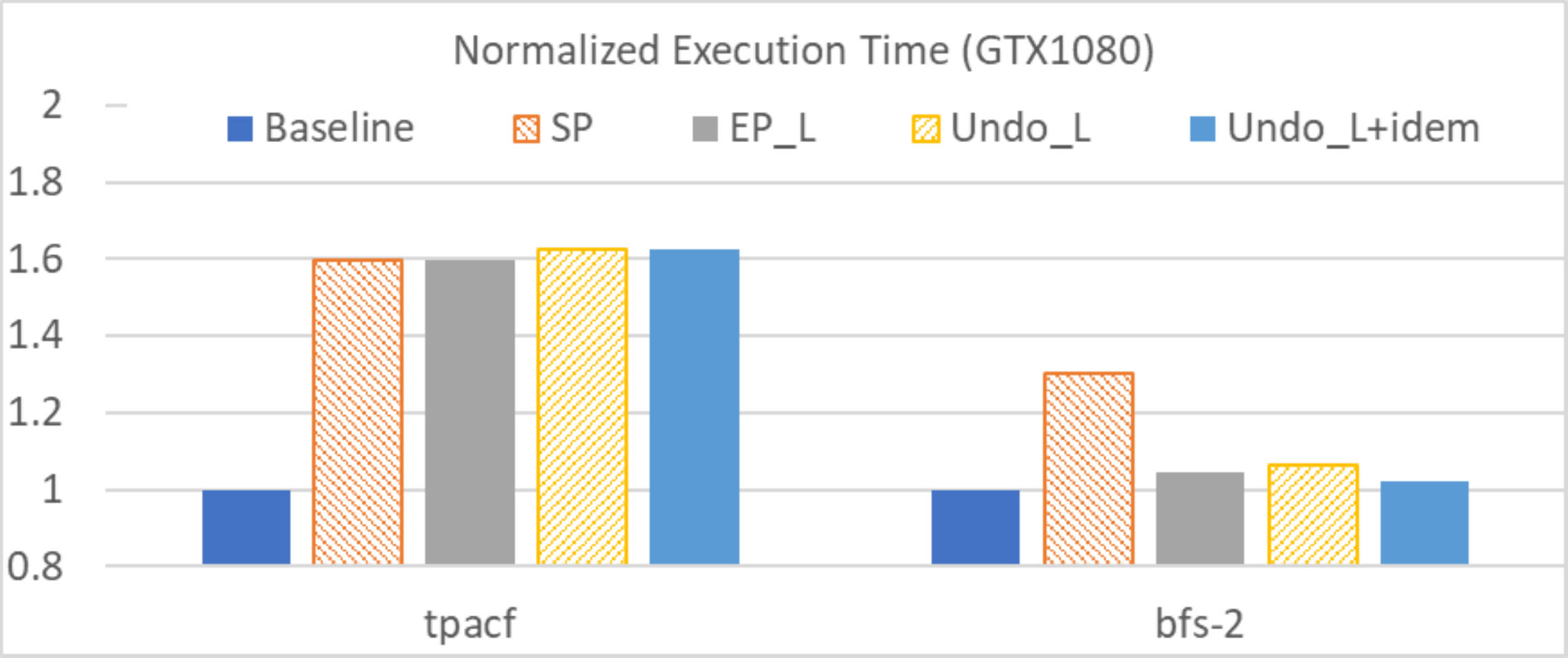}
\caption{Normalized execution time of long-running kernels with long-running CTAs using different persistency models on a GTX1080 GPU.}
\squeezeup
\label{fig:eval_llk_gtx}
\end{figure}

\begin{figure}
\centering
\includegraphics[width=.6\linewidth, trim=2 2 2 2, clip]{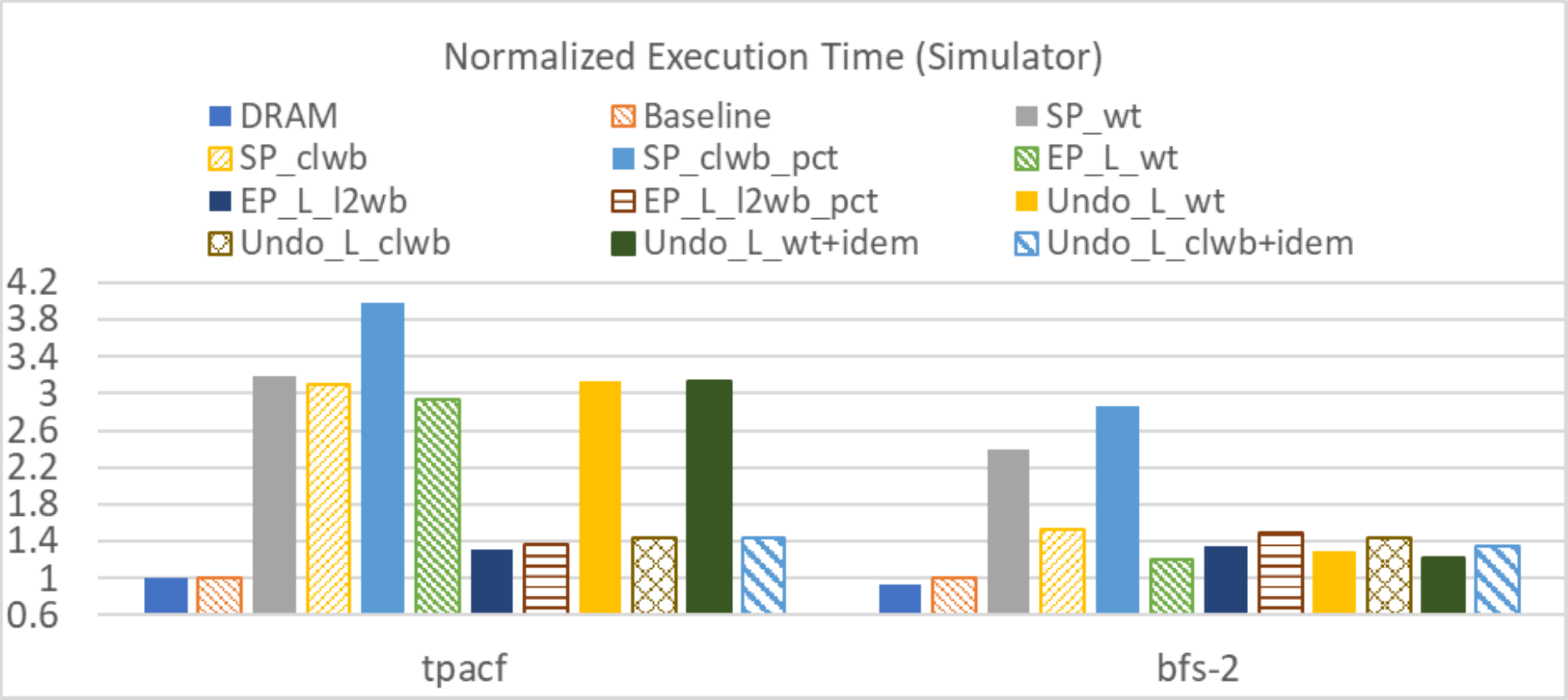}
\caption{Normalized execution time of long-running kernels with long-running CTAs using various persistency models on GPGPUsim.}
\squeezeup
\label{fig:eval_llk_sim}
\end{figure}

Using our criterion in Section \ref{sec:meth}, 2 kernels, tpacf and bfs-2, are classified in the category of long-running kernels with long-running CTAs. The kernel bfs-2 uses atomic operations on global memory variables. So, we choose to use the kernel-level durable transaction model rather than the loop-level model for bfs-2. The performance results on the GTX1080 GPU is shown in Figure \ref{fig:eval_llk_gtx}. As both kernels use shared memory variables, persisting their shadow copies incurs relatively high performance overhead. The loop-level epoch persistency model reduces the overhead for bfs-2 as it enables parallel updates with a single fence at the end of each iteration. For the tpacf kernel, which has higher shared memory usage, such benefit is not clear on GTX1080 but more visible on our simulator (see Figure \ref{fig:eval_llk_sim}). As the shadow copy of the shared memory variables also serves the role of the undo log, the loop-level durable transaction model for tpacf has the similar performance to the loop-level epoch persistency model. The kernel-level durable transaction model for bfs-2 has small overhead as we do not need to back up shared memory at the kernel level.  Since neither kernel is idempotent, the impact of idempotency analysis is limited.

The performance results of the two kernels on the simulator are shown in Figure \ref{fig:eval_llk_sim}.
We can see that the loop-level epoch persistency models have better performance than the strict persistency models. As tpacf uses a high amount of shared memory data, the lw2b instruction at the end of the loop (i.e., EP\_L\_l2wb\_pct) achieves better performance than in-place store.wt instructions (i.e., EP\_L\_wt) as it enables more overlapping among the updates. The bfs-2 kernel, however, shows the opposite behavior due to its few shared memory updates.  

\subsection{Recommended Models}

\begin{table}[]
\caption{Recommended persistency and durable transaction models and their performance overheads.}
\small
\centering
\begin{tabular}{|c|c|c|c|c|}
\hline
Bench   & PM Model   & PM Ohd & DT Model      & DT Ohd. \\ \hline \hline
bfs-1   & EP\_K      & 0.8\%  & Undo\_K       & 629.3\% \\ \hline
histo-1 & EP\_K      & 3.7\%  & Undo\_K       & 16.0\%  \\ \hline
histo-2 & EP\_K      & 3.7\%  & Undo\_K       & 4.0\%   \\ \hline
histo-4 & SP+wt      & 0.2\%  & Undo\_K       & 26.5\%  \\ \hline
sad-2   & SP+clwb    & 0.3\%  & Undo\_K       & 4.8\%   \\ \hline
mriq-1  & EP\_K      & 5.6\%  & Undo\_K       & 16.5\%  \\ \hline
grid-5  & EP\_K      & 7.6\%  & Undo\_K       & 41.6\%  \\ \hline
grid-6  & EP\_K      & 3.0\%  & Undo\_K       & 32.6\%  \\ \hline
histo-3 & EP\_C+l2wb & 16.2\% & Undo\_K       & 153.4\% \\ \hline
grid-1  & EP\_C+clwb & 67.6\% & Undo\_C\_clwb & 77.6\%  \\ \hline
cutcp   & EP\_C+wt   & 0.1\%  & Undo\_C\_clwb & 0.5\%   \\ \hline
lbm     & SP+wt      & -0.6\% & Undo\_C\_wt   & 3.9\%   \\ \hline
sad-1   & EP\_C+clwb & 20.6\% & Undo\_C\_clwb & 22.7\%  \\ \hline
spmv    & EP\_C+wt   & -1.6\% & Undo\_C\_wt   & -0.3\%  \\ \hline
sgemm   & EP\_C+wt   & 0.8\%  & Undo\_C\_wt   & 3.3\%   \\ \hline
stencil & EP\_C+wt   & -1.2\% & Undo\_C\_clwb & 6.6\%   \\ \hline
mriq-2  & EP\_C+clwb & 0.0\%  & Undo\_C\_wt   & 0.3\%   \\ \hline
grid-2  & EP\_C+clwb & 0.0\%  & Undo\_C\_clwb & 0.0\%   \\ \hline
grid-3  & EP\_C+clwb & 16.1\% & Undo\_C\_clwb & 71.0\%  \\ \hline
grid-4  & EP\_C+wt   & 1.6\%  & Undo\_C\_wt   & -1.6\%  \\ \hline
tpacf   & EP\_L+l2wb & 30.6\% & Undo\_L\_clwb & 43.8\%  \\ \hline
bfs-2   & EP\_L+wt   & 20.4\% & Undo\_K       & 21.0\%  \\ \hline
Avg.    &            & 8.0\%  &               & 32.3\%  \\ \hline
\end{tabular}
\label{tab:overhead}
\squeezeup
\end{table}

With the kernel classification criteria in Section V, we list in Table \ref{tab:overhead} the recommended memory persistency model and durable transaction model for each kernel as well as the performance overheads. 
The average performance overhead to support the memory persistency models is 8.0\%. To enable durable transaction with undo-logging, the average performance overhead is 32.3\%. Note that the performance overhead is based on the kernel classification criteria in Section V to explore the performance impacts of persistency models and our architecture supports. A coarser epoch/durable transaction would reduce such performance overhead. 

\subsection{Impact on Write Endurance}

\begin{figure}[t]
\centering
\includegraphics[width=.6\linewidth, trim=2 2 2 2, clip]{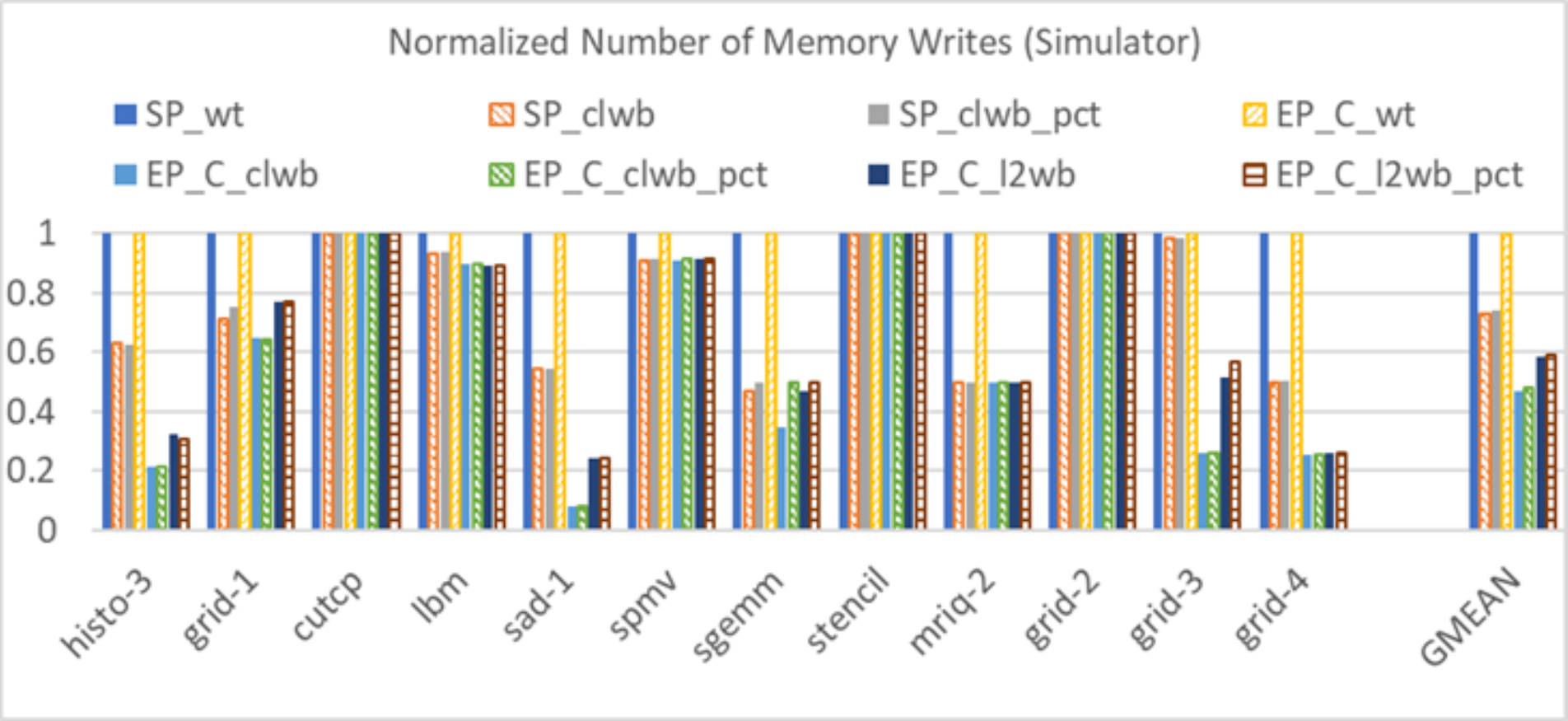}
\caption{Normalized numbers of writes in different memory persistency models for long-running kernels with short-running CTAs.}
\squeezeup
\label{fig:eval_wr}
\end{figure}

Different memory persistency models lead to different numbers of writes to PM, which may affect its write endurance. Here, we use the long-running kernels with short running CTAs to examine this effect. Figure \ref{fig:eval_wr} shows the number of writes for each kernel using different persistency models normalized to the strict persistency model implemented with store.wt instructions. It can be seen that the write-back policy is effective in reducing the write traffic and the CTA-level epoch persistency model further reduces the number of writes by delaying the write backs at the end of the CTAs, which enables more opportunities to combine the updates.

\subsection{Summary}
The key results from our experiments include (1) the strict persistency model incurs higher performance overhead than the epoch persistency models; (2) among the epoch persistency models, epochs with coarser granularities have lower performance overheads and more opportunities to reduce the number of writes; (3) write-through is a good fit for strict persistency while the epoch persistency models work better with clwb; (4) undo logging may introduce significant performance overhead, especially when the execution time of a transaction/epoch is low; and (5) idempotency analysis effectively reduces the overhead of undo logging for various scopes of epochs/durable transactions. 
\section{Conclusions}
\label{sec:conclusion}

In this paper, we adapt, re-architect, and optimize CPU persistency models for GPUs. Besides the architectural support for different persistency models, we highlight that the thread hierarchy in the GPU programming model offers intuitive ways to define the scope of an epoch in epoch persistency. Furthermore, these epochs can serve as boundaries of durable transactions, which are supported through undo logging. We propose idempotency analysis to eliminate unnecessary undo logs, and reduce the size of undo logs when the epoch is not idempotent. 

We design a pragma-based compiler approach to facilitate programmers to express the persistent models. Our experiments show that with our proposed architectural support and optimizations, different memory persistency models can be effectively achieved for GPUs to provide various granularities of recoverability at low performance overhead. Our analysis also reveals interesting difference in supporting memory persistency models in GPUs vs. CPUs.

\bibliographystyle{IEEEtran}

\end{document}